\documentstyle[epsfig,times]{aa}

\makeatletter
\def\@cite#1#2{(#1\if@tempswa , #2\fi)}
\def\@citex[#1]#2{\if@filesw\immediate\write\@auxout{\string\citation{#2}}\fi
  \def\@citea{}\@cite{\@for\@citeb:=#2\do
    {\@citea\def\@citea{;\penalty\@m\ }\@ifundefined
       {b@\@citeb}{{\bf ?}\@warning
       {Citation `\@citeb' on page \thepage \space undefined}}%
\hbox{\csname b@\@citeb\endcsname}}}{#1}}
\makeatother

\title{Multifrequency observations of XTE J0421+560/CI Cam in outburst}

\author{F.~Frontera\inst{1,2}
\and M.~Orlandini\inst{1}
\and L.~Amati\inst{1}
\and D.~Dal~Fiume\inst{1}
\and N.~Masetti\inst{1}
\and A.~Orr\inst{3}
\and A.N.~Parmar\inst{3}
\and E.~Brocato\inst{4}
\and G.~Raimondo\inst{4}
\and A.~Piersimoni\inst{4}
\and M.~Tavani\inst{5,6}
\and R.A.~Remillard\inst{7}
}

\institute{
Istituto Tecnologie e Studio Radiazioni Extraterrestri, TeSRE/CNR, Via Gobetti
 101, 40129 Bologna, Italy
\and
Dipartimento di Fisica, Universit\`a di Ferrara, Via Paradiso 11, 44100
 Ferrara, Italy
\and
Astrophysics Division, Space Science Department of ESA, ESTEC, NL--2200 AG
 Noordwijk, The Netherlands
\and
Osservatorio Astronomico di Teramo, Collurania, 64100 Teramo, Italy
\and
Istituto di Fisica Cosmica e Tecnologie Relative, IFCTR/CNR, Via Bassini 15/A,
 20133 Milano, Italy
\and
Columbia Astrophysics Laboratory, Columbia University, New York, NY 10027, USA
\and
Center for Space Research, MIT, 77 Mass Ave, Cambridge, MA 02139, USA
}

\offprints{frontera@tesre.bo.cnr.it}
\date{Received \today; Accepted }
\thesaurus{06          
           (08.02.5;   
            08.09.2;   
            08.14.2;   
            13.25.1;   
            13.25.5)   
}

\begin{document}

\maketitle

\begin{abstract}

We report on two X--ray observations of the transient source \object{XTE
J0421+560} performed by BeppoSAX, and on a series of observations performed by
the 0.7~m Teramo--Normale Telescope. Outburst peak occurrence time and duration
depend on photon energy: the outburst peak is achieved first in the X--ray
band, then in the optical and finally in the radio. An exponential decay law
fits well the X--ray data except in the TOO2 0.5--1.0 keV band, where erratic
time variability is detected. During TOO1 the e-folding time scale decreases
with energy up to $\sim$20 keV, when it achieves a saturation; during TOO2 it
decreases up to $\sim$2 keV and then increases. This change is correlated with
a spectral change, characterized by the onset of a soft ($\la$2~keV)
component in TOO2 (Orr et~al.\ 1998). This component might originate from the
relativistic jets, while the hard component is more likely associated to
processes occurring in the circumstellar matter and/or near the compact object.
Optical observations show that the object appears intrinsically red even during
the outburst. The nature of the compact object is discussed.

\keywords{Stars: binaries: symbiotic -- Stars: individual (XTE J0421+560) --
          Stars: novae -- X--ray: general -- X--rays: stars}

\end{abstract}

\section{Introduction}

\begin{figure}
\epsfig{figure=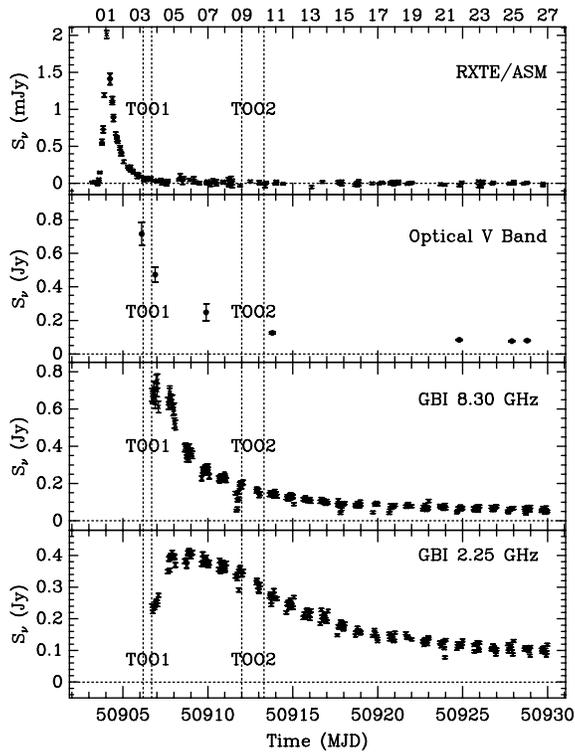,width=\columnwidth}
\bigskip
\caption[]{Light curve of XTE J0421+560 in X--rays (2--12~keV), optical $V$, and radio
(8.3~GHz and 2.25~GHz). Data come from the ASM (upper panel) and GBI (lower
panels) public archives, IAU Circulars and our optical measurements (second
panel). BeppoSAX observations are marked by dashed lines. The upper scale represents
days of April 1998. ASM data after April 12 are upper limits}
\label{asm_gbi}
\end{figure}

The X--ray source XTE J0421+560 was discovered in the 2--12~keV energy band
with the ASM aboard RXTE during a strong outburst, with onset time around 1998
March 31.36, and peak flux of $\sim$2 Crab on 1998 April 1.04 \cite{Smith98}.
Figure~\ref{asm_gbi}, upper panel, shows the light curve of the source on the
basis of the ASM data. The decaying curve after the maximum is consistent with
an exponential law with e-folding time $\tau_{\rm ASM} = 0.860\pm 0.007$~d. The
outburst was also detected in hard X--rays (up to 70~keV) with the BATSE
experiment aboard CGRO at a 20--30~keV peak flux of $\sim$1.1 Crab between 1998
March 31.91 and April 1.04 \cite{Harmon98}. No hard X--ray emission was
observed with BATSE after April 2--3.

Prompt observations at different wavelengths were performed with ground based
telescopes. Observations performed on April 2.63 in the radio band at 1.4~GHz
with the VLA \cite{HM98} showed the presence of a variable source within the
RXTE/PCA error box, whose position was coincident with that of the symbiotic
star \object{CI Cam} (\object{MWC 84}). The light curve of the source obtained
at two different frequencies with the NRAO/NASA GBI telescope \cite{GBI} is
shown in Fig.~\ref{asm_gbi}, lower panels. The decaying curve after the maximum
is consistent with e-folding times $\tau_{8.30} = 3.19\pm 0.08$~d and
$\tau_{2.25} = 12.55\pm 0.08$~d at 8.30 and 2.25~GHz, respectively. The radio
source appeared point-like ($<$0.1$''$ at 22.5~GHz) on 1998 April 3.83 and
became extended on April 5.08. Since April 6.94 it exhibited an almost
symmetrical S-shaped twin-jet \cite{HM98}, strikingly similar to the radio jets
of \object{SS 433}. The jet motion was estimated to be about 26 mas/day,
corresponding to a $0.15~c$ velocity assuming a source distance of one kpc.
This distance was estimated on the basis of absorption considerations and
spectral properties by Chkhikvadze (1970) and Bergner (1995),
\nocite{Chkhikvadze70,Bergner95} and now confirmed with 20\% uncertainty by
Hjellming et~al.\ (1998). \nocite{Hjellming98} We assume this value for
luminosity estimates.

Triggered by the ASM detection of the source, BeppoSAX observed XTE J0421+560
twice (Orlandini et~al.\ 1998). \nocite{Orlandini98} In this {\em Letter} we
concentrate on light curve properties, and on their correlation with the
spectral behavior. In another {\em Letter} Orr et~al.\ (1998) \nocite{Orr98}
attempt to model the source spectrum and its behavior with time. We also report
on optical results obtained in the $V$ and $R$ bands.

\section{X--ray observations and data analysis}

XTE J0421+560 was observed as a Target of Opportunity on 1998 April 3 from
05:03 UT to 17:44 (TOO1), and from 1998 April 9 00:48 to April 10 06:49 UT
(TOO2). These time intervals are indicated in Fig.~\ref{asm_gbi} with dashed
lines.

The BeppoSAX observations were performed with the Narrow Field Instruments
(NFI).  They include a Low Energy Concentrator Spectrometer (LECS, 0.1--10 keV;
Parmar et~al.\ 1997), \nocite{Parmar97} the Medium Energy Concentrators
Spectrometers (MECS, 1.5--10 keV; Boella et~al.\ 1997), \nocite{Boella97} a
High Pressure Gas Scintillator Proportional Counter (HPGSPC, 5--120 keV; Manzo
et~al.\ 1997), \nocite{Manzo97} and a Phoswich Detection System (PDS, 15--300
keV; Frontera et~al.\ 1997). \nocite{Frontera97} Both LECS and MECS have
imaging capabilities, while HPGSPC and PDS are mechanically collimated direct
viewing instruments that use rocking collimators for background monitoring.

We used the default criteria to select good data. Library background blank
field measurements were used for the imaging instruments, while the background
for the collimated instruments was evaluated from offset fields. Data were
analysed using SAXDAS 1.3.0 and XAS 2.1 software packages. From the source
image obtained with LECS and MECS we derived the best source location (epoch
2000): $\alpha$=$04^{\rm h} 19^{\rm m} 46^{\rm s}\!.0$ and $\delta$=$+55^\circ
59'24''$, with an error radius of $50''$ (all the uncertainties reported are
given at the 90\% confidence level). This position is consistent with the RXTE
\cite{Marshall98} and the radio \cite{HM98} position.

\subsection{TOO1}

\begin{figure}
\epsfig{figure=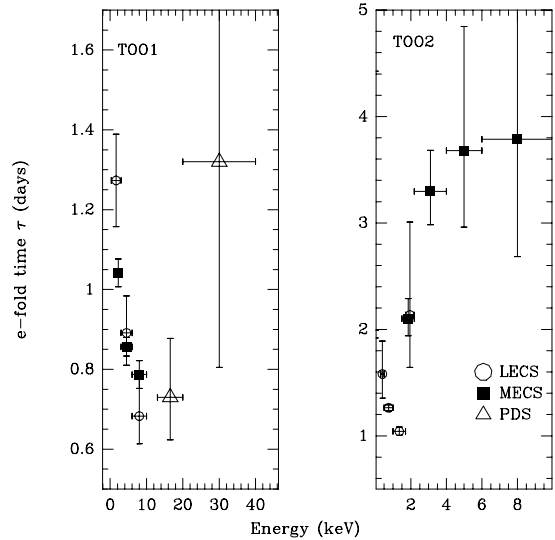,height=8.5cm}
\vspace{-0.7cm}
\caption[]{Dependence of the e-folding time $\tau$ on energy for the two
BeppoSAX XTE J0421+560 observations}
\label{tau}
\end{figure}

TOO1 was carried out during the rising part of the radio flare from CI Cam (see
Fig.~\ref{asm_gbi}) when the source was in a clear optically thick radio state.
It is the first time that a deep X--ray observation is carried out during this
state. The source was detected with all NFI, including PDS. The source spectrum
is very complex, showing a turn-over at $\sim$1.5 keV and a strong Iron line
emission at 6.7 keV \cite{Orr98}. In the 1.5--10~keV the flux level at the
beginning of the observation was $\sim$65 mCrab. To model the decay of the
source we tried simple decaying functions (linear, exponential, power law):
only the exponential law fits well the data. The e-folding time $\tau$ is
$1.27\pm 0.11$~d, $0.92\pm 0.02$~d, and $0.7\pm 0.3$~d in the 0.3--3~keV
(LECS), 1.5--10~keV (MECS), and 15--100~keV (PDS) energy bands, respectively.
The decay constant obtained with MECS is consistent with $\tau_{\rm ASM}$. A
decrease in the decay constant is also evident moving from lower to higher
X--ray energies. The behavior of the e-folding time $\tau (E)$ with energy is
shown in Fig.~\ref{tau}. The erratic time variability of the source on 100~s
time scales is extremely low, once the general trend due to the exponential
decay has been subtracted. The observed variance of $0.077\pm 0.008$~s$^{-2}$
is consistent with that expected assuming a Poissonian noise ($0.071\pm
0.007$~s$^{-2}$). We have also searched for periodic or quasi-periodic
oscillations from the source: there is no evidence for such oscillations in the
frequency band from 10$^{-4}$ to 50~Hz.

\subsection{TOO2}

\begin{figure}
\centerline{\epsfig{figure=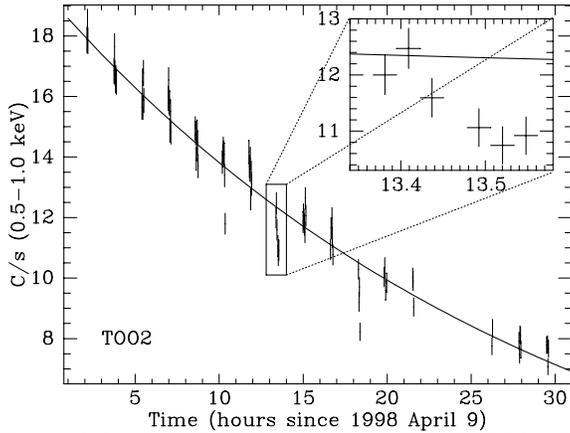,width=\columnwidth,angle=270}}
\vspace{-3cm}
\caption[]{0.5--1.0 keV LECS light curve for TOO2 rebinned at 100~s, together
with the exponential fit to the data ($\chi^2_\nu = 3.71$ for 80 degrees of
freedom). The inset shows an enlargement of the boxed data}
\label{lecs_curve}
\end{figure}

TOO2 was carried out during the decay part of the radio flare, during the
optically thin phase of radio emission. The source was detected with the LECS
and the MECS, but not with the PDS. In the 1.5--10~keV, the flux level at the
beginning of the observation was $\sim$8 mCrab, while the 2$\sigma$ upper limit
to the flux in the 15--40~keV band was 0.5~mCrab. An exponential decay law fits
well the data but in the 0.5--1.0 keV band (see Fig.~\ref{lecs_curve}), where
the source spectrum shows the emergence of a soft ($\la$2 keV) component
\cite{Orr98}. The e-folding time is $1.2\pm 0.2$~d and $2.3\pm 0.3$~d in the
0.3--3~keV (LECS) and 1.5--10~keV (MECS) energy bands, respectively. The latter
value is about 3 times larger than $\tau_{\rm ASM}$, indicating that the source
decay slows down. A possible explanation could be the presence of a steady
component. The decay constant has a different behavior as a function of energy
with respect to TOO1, as is shown in Fig.~\ref{tau}: the e-folding time
decreases up to about 1~keV and then increases with energy. The variance of the
1.5--10 keV data, once the decaying law has been subtracted, is completely
consistent with a Poissonian statistics, while the 0.5--1.0 keV light curve
shows significant variability on a 100~s time scales (see
Fig.~\ref{lecs_curve}).

\section{Optical observations and data analysis}

Optical photometry in $V$ and $R$ bands has been performed on 1998 April 6.9, 
10.8, 21.8, 24.9 and 25.9 with the 0.70~m TNT (Teramo--Normale Telescope) of 
the Teramo Observatory, equipped with a Tektronics TK512CB1-1 CCD camera (pixel
size of 0.46 arcsec/pixel). A total of 37 frames were collected, with exposure
times ranging from 60 to 600~sec for the $V$ band and from 40 to 200~sec for
the $R$ band, depending on the seeing conditions. The frames were debiased and
flat-fielded in the usual way, reduced with simple aperture photometry inside
MIDAS, and calibrated using field stars \cite{Granslo98}. The uncertainty on
our data is mainly due to calibration errors.

By combining our photometrical data with those published by Garcia et~al.\
(1998) \nocite{Garcia98} and Hynes et~al.\ (1998) \nocite{Hynes98} we find that
the $V$ magnitude of the star faded from $9.25\pm 0.1$ mag on April 3.13 to
$9.7\pm 0.1$ on April 3.87, to $11.2\pm 0.05$ on April 10.80. The light curve
of the source in the $V$ band is shown in  Fig.~\ref{asm_gbi}, second panel. No
remarkable changes in the $V$--$R$ color are noticed during the outburst: it
remains $\sim$1.1. As also noted by Hynes et~al.\ (1998), this is the same
value measured by Bergner et~al.\ (1995) \nocite{Bergner95} during quiescence.
Then, from April 21 the magnitude varies around $V\sim$11.8: this is $\sim$0.2
mag fainter than the mean value reported by Bergner et~al.\ (1995).
\nocite{Bergner95} Appreciable flickering with amplitude $\sim$0.3 mag on a
timescale of about 1~hr is seen on April 6, while no variations are detected in
the later observations. Assuming an exponential decay also for the optical
light curve, and once the contribution of the quiescent optical counterpart has
been subtracted, we find that $V$ and $R$ e-folding times are $\tau_V =
1.6\pm0.7$~d and $\tau_R = 1.19\pm0.14$~d for TOO1; $\tau_V = 3.54\pm0.37$ d
and $\tau_R = 3.16\pm0.35$~d for TOO2. These values are compatible with those
obtained from the BeppoSAX observations.

\section {Discussion}

The observed outburst shows several peculiarities, some of which could give
hints on the nature of the source. The X--ray outburst is followed by a radio
outburst and by a change in the brightness of the optical counterpart (CI Cam)
associated with XTE J0421+560.

Outburst peak occurrence time and outburst duration depend on photon energy. It
is evident from Fig.~\ref{asm_gbi} that the outburst peak is achieved first in
the X--ray band (April 1.04), and eventually at 2.25~GHz ($\sim$April 6), with
the X--ray outburst duration much shorter than the radio one. This can be an
indication of a transition from an optically thick to an optically thin medium.
It is worth noting that this transition coincides with the onset of the radio
jets.

The X--ray e-folding time depends on photon energy. During TOO1 it appears to
decrease with energy up to about 20~keV where it achieves a saturation; during
TOO2 it decreases up to about 1~keV and then increases. One of the major
changes occurred in the energy spectrum from TOO1 to TOO2 is the emergence
during TOO2 of the soft component below 2 keV, modeled with two narrow emission
lines with energies at $\sim$0.7 and $\sim$1.1 keV  that smoothly decrease
during the observation \cite{Orr98}. From Fig.~\ref{tau} we can see that this
soft component is characterized by a different temporal behavior: it evolves
more rapidly than the corresponding hard ($\ga$2 keV) component. A possible
explanation for this behavior is the association of this component to the radio
jets (or to the process that created them). This is supported by the presence,
{\em only} in the soft component, of 100~s time scale temporal variability (see
Fig.~\ref{lecs_curve}), implying an upper limit for the emitting region of
$3\times 10^{12}$~cm. The hard component, also because of the presence of
various emission lines \cite{Orr98}, could be associated to the circumstellar
matter around the XTE J0421+560/CI Cam system.

The similarity between the hard X--ray and optical e-folding time scales should
suggest a common nature for these emissions, such as reprocessing of X--rays
from the central object into optical light. However, other optical properties
are not in agreement with this interpretation. In the optical band, where most
of the outburst flux is emitted in the red part of the spectrum, the $V$--$R$
color does not change appreciably during the decay to quiescence. This is
surprising because the outburst would imply a heating of emission zones such as
a disk or the side of the mass-losing star facing the compact object. For
instance, cataclysmic binaries and X--ray novae (XN) become appreciably bluer
during the outburst and most of their luminosity is emitted at low wavelengths
in the optical (see, {\em e.g.\/}, van~Paradijs \& McClintock 1995).
\nocite{vPmC95}

The presence of an optical and a radio outburst associated with the X--ray one
is typical of XN outbursts in which the compact object is a neutron star (NS)
or a black hole (BH) (see review by Tanaka \& Shibazaki 1996). \nocite{TS96}
Also the presence of relativistic jets of matter has been considered typical of
NS/BH system. In spite of this, the features shown by XTE J0421+560 rise
several doubts on a similar nature of the compact object.

\begin{figure}
\epsfig{figure=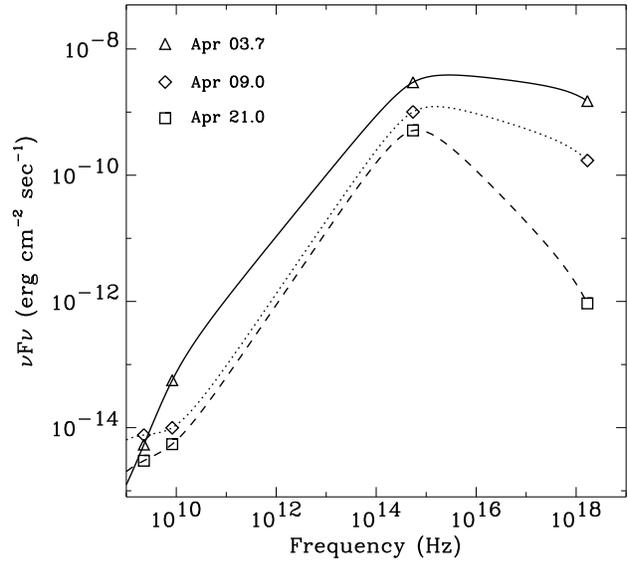,width=\columnwidth}
\vspace{-0.8cm}
\caption[]{$\nu F\nu$ plot of XTE J0421+560 at  different times during outburst.
The lines are the interpolation with splines of the data points. The optical
point is the average of $B, V, R$ measurements. The Apr 21 X--ray point
is estimated from the TOO2 decay law}
\label{vFv}
\end{figure}

The 2--12~keV X--ray luminosity at the maximum of the outburst, assuming an
isotropic emission of the radiation, results to be $5.1\times 10^{36}$
erg~s$^{-1}$. For comparison, known NS or BH XN exhibit higher peak
luminosities \cite{TS96,TL95}. The X--ray energy emitted in the outburst by XTE
J0421+560 is estimated to be $3.7\times 10^{41}$~erg in the X--ray band, $1.6
\times 10^{41}$~erg in the optical band ($R$ filter), $4.6 \times 10^{36}$~erg
at 8.3~GHz and $1.8\times 10^{36}$~erg at 2.25~GHz. For comparison that emitted
by NS and BH XN in the X--ray band is much higher ($10^{43}$--$10^{45}\,$erg;
Tanaka \& Shibazaki 1996). \nocite{TS96} We also evaluated the bolometric
energy emitted in the outburst, by making use of the X--ray, optical and the
radio data. The derived broad band logarithmic power per photon energy decade
(the $\nu F\nu$ spectrum) is shown in Fig.~\ref{vFv} at different times during
outburst.  The upper limit to the total energy emitted during the entire
outburst is 1.5$\times 10^{44}\,$erg, similar to that estimated for classical
novae (CN) systems, although X--ray and $\gamma$-ray emission is generally not
detected in them \cite{Warner95}.

This object seems to have experienced an impulsive eruption, with a
characteristic time scale more similar to that of a dwarf nova outburst
($\sim$1 week) than to those of CN or XN ($\sim$months), which also show more
complex light curve features \cite{TS96}. Furthermore, the X--ray emission from
XTE J0421+560 does not show the erratic flux variability of NS or BH XN
systems, that is likely associated to the accretion process that takes place in
these systems.

In conclusion, X--ray observations clearly indicate the presence of a compact
object in XTE J0421+560, but the available data do not allow its nature to be
pinpointed. The presence of X--ray, optical and radio outburst, together with
relativistic jets, is typical of neutron star and black hole systems. On the
other hand, the temporal behavior and the energetics are compatible also with a
white dwarf system. Our data suggest that the soft X--ray emission might
originate from the relativistic jets, while the hard component is more likely
to be associated to processes occurring in the circumstellar matter and/or near
the compact object.

\begin{acknowledgements}
This research has made use of data obtained from the NRAO/NASA GBI public
archives. BeppoSAX is a joint Italian and Dutch programme. This research was
supported in part by the Italian Space Agency. We wish to thank Bob Hjellming
for providing in advance his new result on the source distance.
\end{acknowledgements}

\end{document}